\documentclass[8pt,draft,notref,notcite]{article}

\usepackage{color,longtable}
\usepackage{amsmath,amssymb,graphicx}
\usepackage{amssymb,graphicx}

\newcommand{\beq}{\begin{equation}}
\newcommand{\eeq}{\end{equation}}

\newcommand{\be}{\begin{eqnarray}}
\newcommand{\ee}{\end{eqnarray}}

\usepackage{amsfonts,amsmath}
\usepackage{latexsym}

\def\cN{{\cal N}}

\def\R{{\mathbb{R}}}

\def\cN{{\cal N}}

\def\N8{{\cN\!=\!8}}

\begin{document}

\begin{center}
{\Large \bf An exceptional story: Symmetries and dualities between Maximal supergravity and General relativity$^*$}\\[7mm]

-  Bernard L. Julia  -
 \\[4mm]
{\sl Laboratoire de Physique de l'\'Ecole Normale Sup\'erieure \\ 
CNRS, ENS, Universit\'e PSL,  Sorbonne Universit\'e, \\
Universit\'e de Paris . F-75005 Paris, France\\[17mm]
	}
\vspace*{0.8cm}

\begin{minipage}{12cm}\footnotesize
\textbf{Abstract:} We present the historical path from General relativity to the construction of  Maximal  $\mathcal{N}_4 = 8$ Supergravity
with a detour in D=10 and 11 dimensions. The supergravities obtained by toric dimensional reduction and/or by reducing the number of supersymmetry 
generators have large exceptional duality symmetry groups and exhibit a remarkably uniform pattern across all values of  $\mathcal{N}_D$ and D.
In particular (bosonic) General relativity  fits in as the simplest case and anchors us to the Real world.
Dimensional reduction to 2 dimensions brings us to affine Kac-Moody groups and their semi-direct products with a real form of the Witt algebra: there is "integrable Magics". 
Integrability of 4D Gravity and of its reduction to 2D is considered with their "Twisted self-duality".  
Hyperbolic Kac-Moody symmetries appear after reduction to 1D: this leads to "chaotic Magics".  
We then discover "Borcherds"-Kac-Moody symmetries that allow us to rewrite in any dimension all matter equations of motion as Twisted self-duality: "Algebraic geometric Magics". 
Finally a "BF" metasymmetry $\Sigma$ exchanges negative quartets of Fermionic dimensions with Bosonic ones inside two Magic triangles.
A third ubiquitous triangle of symmetries from Invariant theory resists unification despite its strong resemblance to the others.
The prospective remarks include seven Challenges.
\end{minipage}\end{center}

\vspace{2cm}

\noindent
*Invited contribution to the book {\it Half a Century of Supergravity}, eds. A. Ceresole and G. Dall'Agata 
      \newpage

\tableofcontents

\section{Introduction. Classical/quantum duality.}

When in the fall of 1974 I constructed dyon solutions of the classical Yang-Mills-Higgs (Georgi-Glashow) D=4 field theory equations at Princeton university I became fascinated by \textit{duality} and especially by the Dirac-Schwinger-Zwanziger quantization condition that is expressed in terms of the invariant defining the electric magnetic (quantum) duality group $SL(2,\mathbb{Z})$. Spin=eg’-e’g=$nh^{Planck}\slash 4\pi$ where (e,g) and (e’,g’) are the electric and magnetic charges of two dyons.

Particularly exciting to me was the possibility to generate spin in 4D, especially a spin one half from boson fields only, that was against general wisdom. The Coleman-Mandelstam fermionization (in the D=2 case) was discovered a few months later. Another curious remark in my paper with Tony Zee was the similarity of the Higgs field with a euclidean time component of the gauge potential: this was used by M.K. Prasad and Charles Sommerfield and then by Eugen Bogomolny in the BPS bound that generalizes the instanton inequality (1975-6). Still in 1975 in Princeton Curtis Callan mentioned to me a question of Sidney Coleman about our observation of the non-discreteness of the dyon electric charge at the classical field theory level. I answered quickly that \textit{the classical limit is not unique}. This fact is a key property of our quantum world, but it has not been systematically incorporated in standard textbooks. It probably has not even been fully exploited. It may help to think about different limits of vanishing (adimensionalized) Planck's constant, each of which must be specified by a choice of the other parameters to be kept constant. In the full quantum description spin is quantized in multiples of $h^{Planck}\slash 4\pi$, electric charges are multiples of the smallest particle value $e_{min}$ and if we neglect CP violation (topological) magnetic charges are integer multiples of the classical (electrical) field theory value eg. $h^{Planck}\slash (4\pi e_{min})$.
Hence I propose the 
	
\textbf{First Problem}:
In a situation with several possible classical limits could one measure for each one of them one \textit{degree of classicality} of the configuration or of the process or several such degrees?  Could one then determine in each case 
which one of these semi-classical approximations is most appropriate (e.g. for particle-wave complementarity or  for electromagnetic duality between conducting and superconducting phases)? An electric particle description is equivalent to a magnetic wave description and conversely.

\section{From Strings to $4D$ SUSY and gravitation}

Let me backtrack to 1972 in order to present the field of view at that time. The string picture of dual models had already been developed by Yoichiro Nambu and others since 1969. LPT my homelab in Orsay was close to the CERN theory division so I had followed the discovery and literature on 4D $\mathcal{N}_4 = 1$ global supersymmetry. $\mathcal{N}_D$ will be the number of spinors of Susy charges in dimension D. As I was quite impressed by my seniors Andr\'e Neveu and Jo\" el Scherk I also kept an eye on dual string models during the period 1972-74. In 1972 with Andr\' e Jo\"el developed further his zero slope limit idea (that had answered a question of Roland Omn\`es) in order to encompass Yang-Mills field theory. In 1973 Tamiaki Yoneya found a relation between General Relativity and the small slope limit of closed string theory which was transmuted into the revolution of preonic string theory by Jo\"el and John Schwarz in April 1974. This sequence of events might be called \textit{the String resurrection}. 
	
Jo\"el had suggested that I should read his review of the dual string model (to be published in 1975). In the fall of 1974, enlightened by these ideas and puzzled by the (globally) supersymmetric models of Julius Wess and Bruno Zumino, I received a fellowship from Princeton University and crossed the Atlantic ocean on the liner France with Thibault Damour on board and Jo\"el’s review in my suitcase. I was to recommend studying the latter to Edward Witten among others. 

In Princeton I studied the 1975 preprint by Jo\"el and John on a  ‘Dual field theory of quarks and gluons’ and then I studied old papers on 5 and 6 dimensional unified or projective field theories in particular one by Wolfgang Pauli himself involving spinors. I finally convinced myself despite friendly warning and historical advice by Valentine Bargmann see for instance \cite{Halpern2004} that one or two \textit{extra space dimensions} were real  in some sense\footnote{e.g. meaning one could localize things in these extra-dimensions or the existence of (Kaluza-)Klein 4D modes.}. In those days L. Faddeev, S. Coleman, most people… even Y. Nambu in the early 1970's did not believe in physical extra-dimensions. To the list one may add Paul A. M. Dirac who refused  in Trieste around 1980 to answer my only question to him ever. More precisely he postponed it by a devastating “Maybe later.” so I have renamed it

\textbf{Second Problem}:
“[Professor Dirac] should we take the fifth and sixth tangent dimensions in the 1928 Dirac electron equation as physical (à la Kaluza-Klein) or are they merely mathematical traces of (broken $SO(2,4)$) \textit{conformal invariance} in 4D as used in [your] the 1936 wave equations in conformal space?” 

In 6D scale invariance makes the five dimensional light cone four dimensional. My preference for the first answer came from the 6D symmetry of the 4D Clifford algebra. In 1977 It was 
clear to me that dilations are the imaginary counterpart in $SO(2,4)$ to $x^5 \,  x^6$ (chiral) rotations in $SO(3,3)$ or $SO(1,5)$ , see section {\bf 5}. Since at least 1982 (answering a question after a talk in Moscow) I expressed publicly my subjective preference for a (3,3) signature, we shall return to this as a motivation for our last Challenge in section {\bf 7}.
\bigskip

 The suggestion of extra dimensions came from classical unified gravity theories (Theodor Kaluza…) but another more specific motivation came from quantum string theory (Claud Lovelace, 26D, 10D,…). I had the privilege to be able to work for a couple of years on the first idea without too much competition, at first probably only Murray Gell-Mann and Yuval Ne’eman took physical extra dimensions seriously. I realized that in scalar gravity theories à la Kaluza it is more predictive and beautiful to implement all the five dimensional equations of motion. After returning to France I convinced Jo\"el that this applied to what was subsequently called spontaneous compactification. 

Among other activities in Princeton I tried to participate in the semi-classical efforts of Roger Dashen, Brossl Hasslacher and Andr\' e and learned about \textit{integrability}. I also missed a chance to cosign a nice paper with Tullio Regge and Fernando Lund because I wanted to concentrate on the “more important” problem of \textit{fermionization in dimensions higher than 2}. 
This was one year before the papers by R. Ward on the self-dual Yang-Mills equations and those of C.N. Yang or A. Belavin and V. Zakharov on their inverse scattering form which date back to early 1977. One can also remark that both Mikio Sato’s work and B. Kostant’s progress with western colleagues on integrable systems started that same year.  

The string community had recovered from its failure to incorporate the quark-parton model and joined the supersymmetry community. It then slowly reexpanded with the new (super)String program. In late 1976 Andr\' e mentioned to me some work on open string \textit{modular invariance} by an impressive German working at CERN: Werner Nahm who had just introduced helicity partition functions. Werner was a close friend of top mathematicians Egbert Brieskorn and Don Zagier. 
In Princeton the influence of Tullio (himself a confidant of Enrico Bombieri), Arthur Wightman or Elliott Lieb and others like Valentine, Eugene Wigner, John Wheeler… contributed to the flourishing of our field. One should also remember that C.N. Yang, Y. Nambu and many other physicists had a high interest in modern Mathematics.

In the West \textit{supersymmetry} both in two and in higher dimensions emerged from dual string studies. The D=2 superconformal algebras for the Pierre Ramond and Neveu-Schwarz sectors of the eponymous superstring model had been constructed in 1971. A field theory realization was worked out soon thereafter by Jean-Loup Gervais and Bunji Sakita who reproduced the operator result in a superconformal path integral framework with 2D supergravity invariance group (ie local SUSY) partially fixed. Only in September 1976 was the full reparameterization invariance restored by Lars Brink, Paolo Di Vecchia and Paul Howe and one day later by Stanley Deser and Bruno. The latter pair had simplified in April the $\mathcal{N}_4 = 1$ supergravity action constructed in March partly at ENS by Peter van Nieuwenhuizen, Sergio Ferrara and Dan Freedman. These were intense days and nights of work by a very dedicated and friendly but competitive community. The rewriting of the Dual model amplitudes as functional integrals had begun with the bosonic case in March 1970.

For a while Bruno (and Julius) had been proposing the introduction of a spin 3/2 Rarita-Schwinger field (superconnection) and Stanley had explored systematically perturbative corrections in quantum gravity theories with matter, all in 4D. Stanley was also a strong advocate of a \textit{deformation} method to gauge global symmetries (actually it makes an abelian gauge theory nonabelian while preserving a global symmetry, for this view see \cite{Julia1986}). This is called the (Emmy) Noether method, it was used to construct the first supergravity actions and is still indispensable. 

One now thinks of supergravity as the massless sector or as low energy approximation of a superstring model  and it is used as a tool for studying  \textit{superstring} theories. \textit{Supergravity}’s intrinsic motivation was the ‘gauging‘ of global supersymmetries with possible applications to \textit{unification of forces and regularization of quantum divergences in gravity theories}. At the beginning of 1977 I returned to LPT-(ENS Paris by then) where I had the privilege to witness year II of Supergravity and the beginning of the construction of 4D extended supergravities.

\section{$D=4, \mathcal{N} _4 \leq 4$ SUGRAS. The iNdex $\mathcal{N}$  }

The Poincar\'e-supergravity equations generalize the Einstein-Hilbert ones and are on-shell invariant under (Lie) superalgebras that  the Poincar\'e Lie algebra. The odd generators of (extended) $\mathcal{N}_4$-supersymmetry in 4D are $\mathcal{N}_4$ Majorana 4-component spinors which transform as a vector of $SO(\mathcal{N}_4)$. In $\mathcal{N}_4$=1 supersymmetry three invariance groups are acting together: axial or \textit{chiral/dual} transformations, supersymmetry and translation invariance. One consequence is that the corresponding conserved Noether currents form a supersymmetry (coadjoint) multiplet as explained by Sergio and Bruno at the end of 1974. Specifically in supersymmetric models duality symmetries act as chiral transformations on the spinor fields and as (Hodge) dualities on the vector fields. 

The discovery of the duality (electric-magnetic) invariance of Maxwell’s equations by Heaviside-Larmor dates back to 1892-1897. The classification of the symmetries of the S-matrix for globally supersymmetric theories by Rudolf Haag, Jan Lopuszanski and Martin Sohnius was a breakthrough that took place eighty years later in 1975. Their theorem established the possibility of  \textit{internal} $U(\mathcal{N})$ \textit{duality groups} of symmetries in the conformally invariant case assuming no infrared difficulties and a finite number of types of particles. It may teach us a lesson of patience. When $\mathcal{N}=4$ SU(4) may replace the symmetry U(4). The mixed case was not treated but the purely massive situation allows only internal central charges to appear in the anticommutators of spinorial charges. The internal symmetry is actually more restricted as shown by Werner by bounding the spin or by assuming simplicity of the global Susy algebra see next section \textbf{4}. The complement of $O(\mathcal{N})$ inside $U(\mathcal{N})$ internal symmetry acts on the odd supergenerators of supersymmetry with (non-abelian) chiral transformations. $U(\mathcal{N})$ is called \textit{R-symmetry}. For $\mathcal{N}=1$ R=U(1) is exactly the group of chiral transformations when the field of angles is the field of real numbers. 

At the end of 1976 Sergio, Jo\"el and Bruno verified the global duality invariance $U(\mathcal{N})$ of $\mathcal{N}$=2 and 3 pure supergravities equations of motion. The case of Poincaré supergravity goes beyond the global supersymmetry theorem as it is not conformally invariant and assumes local supersymmetry. The trio still conjectured an R-algebra  $su(4) \sim so(6)$ of off-shell invariances of 4D $\mathcal{N}_4 = 4$-supergravity as expected from the so(6) symmetry of a “toric reduction” from 10 (stringy) dimensions. This led to the idea that there should exist two dual field realizations of the $\mathcal{N}_4=4$ theory with SO(4) respectively SU(4) off-shell duality symmetry, see the following section. For $\mathcal{N}_4 = 3$ U(3) duality should be the obvious subgroup of SU(4). 

In 1977 Eug\`ene Cremmer replaced Bruno in the previous trio to construct the first few orders of the SO(4) theory that is non-polynomial in the scalar field; this was also done by Ashok Das. They then went on to assume its U(4) on-shell symmetry in order to constrain further the complete action. In a second paper they followed up the off-shell invariant SU(4) invariance idea mentioned above. The point is that under the SO(4) of the first $\mathcal{N}_4=4$ supergravity the supercharges form a quadruplet (a vector) whereas under the Spin(4) in the SU(4) coming from 10 dimensions the supercharges transform as a spinor. When they constructed the new SU(4) action a big present was waiting for them namely they not only found the expected U(1) duality needed to complete U(4) but they discovered a full \textit{non-compact and hidden SU(1,1) group of on-shell duality symmetries} (Christmas 1977). 

After lunch one day I had to reassure Andr\'e who at the time was explaining some SU(1,1) properties to the second trio: my point was that Elie Cartan’s theorem on the indefiniteness of the Killing form of \textit{non-compact semi-simple Lie groups} was compatible with unitarity once projected to the symmetric space constructed by quotienting out the maximal compact subgroup U(1) of SU(1,1) to describe the scalar manifold, this will be a general feature of dualities. By then I was getting aspired into the extended supergravity twister.

In higher $\mathcal{N}$-supergravities SU(1,1) becomes a larger and even more surprising non-compact duality group. This was published in September 1978 after the reductions of 11D supergravity to lower dimensions along compact tori \cite{CJ1978},  see section \textbf{5}. Three years later Bruno and Mary K. Gaillard would establish that duality symmetries in 4D have to lie within a maximal (non-compact) group $Sp(2n,\mathbb{R})$, with n the number of propagating vector fields. Indeed $sp(2,\mathbb{R})=sl(2,\mathbb{R})=su(1,1)$ and  the duality symmetry group, split $E_7$,  of our $\mathcal{N}_4 =8 $ theory with 28 vector potentials is strictly included in $Sp(56,\mathbb{R})$. The \textit{doubling of the vector potentials} \cite{CJ1978} by adding their magnetic duals allows to unify their equations of motion and Bianchi identities this will be discussed in section \textbf{6}. 

The 1978-79 \textit{d\'etour} via eleven dimensions we are about to present could be qualified as a mathematical trick by 4-dimensionists. But as I argued extra dimensions might be physical. Still today the verification of higher duality symmetries assumes their existence. It uses all the available information like the superstring model’s 10 dimensions for the construction of $\mathcal{N}_4 = 4$ SU(4) theory. A complete deduction of all these dualities from first principles escapes us. In particular no fully geometrical derivation of $\mathcal{N}_4 = 8$ supergravity is known yet. 

A \textit{toric reduction} is the compactification of k space dimensions on a k-torus (k=11-D if one starts in $11D$) followed by (consistent) truncation to the homogeneous sector with all fields independent on k Killing coordinates. For any dimension D we shall \textit{define the iNdex $\mathcal{N}$} of a $\mathcal{N}_D$ supergravity to be equal generically to $\mathcal{N}_4$ the number of spinors of supersymmetry of the related \textit{pure} supergravity in 4D (either its toric reduction to 4D or inversely the 4D theory of which it is the toric reduction: its 4D \textit{"oxidation"}). We think of redox as \textit{vertical moves}. There is one exception: there is no D=4 $\mathcal{N}_4 = 7$ theory because CPT adds 1 extra fermion of supersymmetry generators so it is nothing but the CPT self-dual $\mathcal{N}_4 = 8$ theory. For a better regularity we shall henceforth define the iNdex for the maximal supergravities (seventh) column to be $\mathcal{N}= 7$ but one keeps $\mathcal{N} = \mathcal{N}_4$ for the others. It turns out that the columns  of the \textit{pure} D=4  $\mathcal{N}_4$-supergravities play a very special role in any D. 

We shall see in the Appendix-section \textbf{8}  (for the first triangle/trapezoid the so-called Magic SUGRA-triangle see \cite{Julia1980}) that the $B:=(D-2) \le 9, F:=(8 - \mathcal{N}) \le 8$ positive \textit{BF=BoseFermi-quadrant} contains 2 beautiful and surprising “triangles” of theories. Contrary to the SUGRA triangle of physical supergravities the second triangle so-called  Magic SPLIT-triangle is obtained by relaxing supersymmetry, we shall discuss it in section 7. Changes of $\mathcal{N}$ at constant D like Susy truncations are called \textit{horizontal moves}. Let us define \textit{$D_{max}$} the highest dimension to which a given supergravity can be oxidized, it depends on $\mathcal{N}$.   
Murray who was leading the hunt proved that $\mathcal{N}_4$ helicity lowering generators acting in short representations could respect the (finite component) field theory bound of maximal helicity equal to 2 only for $\mathcal{N}_4 \le 8$. The $\mathcal{N}_4 = 8$ maximal case is self-dual and irreducible with respect to helicity reversal or CPT. A similar helicity argument implies that any $\mathcal{N}_4 > 4$ theory has to be a supergravity with a graviton. This uses the absence of any interacting theory with maximal spin 3/2. In other words the maximal supersymmetric theory with no spin higher than 1 has $\mathcal{N}_4 = 4$. In March 1978 most supergravity pioneers were still solidly anchored to 4D. For instance the most advanced attempt by Bernard de Wit and Dan in July 1977 at constructing the nonpolynomial $\mathcal{N}_4 = 8$  supergravity assumed correctly SU(8) duality invariance but it used only $D=4$ methods and hit a wall of complexity. 

One important open question is to compare and contrast the status of the two CPT self-dual and \textit{maximal} theories: $\mathcal{N}_4 = 8$ supergravity and $\mathcal{N}_4 = 4$  super Yang-Mills. Whereas the latter is conformally invariant even at the quantum level, the quantum finiteness of the former is still almost universally disbelieved. Let us jump ahead this time by four years: in July 1982 I placed a bet against Michael Green with Marcus Grisaru as referee. My deliberately rather general formulation was the
	
Bet  (or \textbf{Third Problem}):
“(Prove that) Symmetries - in particular duality symmetries and their possible fermionic extensions or unavoidable higher symmetries to be discovered - will guarantee \textit{perturbative UV finiteness} of $\mathcal{N}_4 = 8$  pure supergravity (in 4D) and not only that of superstrings of type II”.

 I won a battle against the false quasi-universal claim of three loop 4D divergences but the search for true divergences is still ongoing thanks to the amazing work of Zvi Bern, Lance Dixon, David Kosower and their many collaborators. Other difficult issues remain undecided like UV completeness. The nonperturbative version of the bet would require the inclusion of all the solitons and higher branes in particular membranes for which the consensus is that there is a unique theory (\cite{Witten1995} see also \cite{HT1994}). Hence no non perturbative bet should be considered. Superstring theory has a high degree of unicity (up to gauging deformations) but we are lost in its solution space. 

\section{The $D=10-11$ detour, $\mathcal{N} =7$ and first magics}

Higher spacetime dimensions were expected if string theory was to be relevant at all. The number of components of one higher dimensional spinor is 32 in 11D (Majorana)  or 16 in 10D (Majorana-Weyl spinor) but $32=8\times 4$ is the number of charges of maximal supergravity in 4D. These are numbers of fermionic dimensions. There are many ways to organize those: 32 (odd) supercharges form either  $\mathcal{N}_4 = 8 $ quartets or $\mathcal{N}_3 = 16 $ doublets etc. It is in Europe -the birthplace of string models- that higher dimensions came finally under close scrutiny from the two sides of the Rhine river.

Firstly in a landmark paper on open and closed superstrings Ferdinando Gliozzi, David (Olive) and Jo\"el discovered in September 1976 \textit{ambient supersymmetry} in 10 dimensions after some prompting by Victor Ogievetsky. Four months later they proposed that SU(4)=Spin(6) supergravity could appear after reduction to 4 dimensions and this was crucial for the 1977 $\mathcal{N}_4 = 4$ supergravity construction. In December 1976 a nice result followed: Lars, Jo\"el and John determined that $\mathcal{N}_D  =1$ pure super Yang-Mills theories lived only in D=2-3?, 4, 6 or 10 (actually in 3D and not 2 Kugo-Townsend 1983). 

The second set of observations came from Werner in the summer 1977. He classified field representations of supersymmetry algebras: Poincar\'e (non simple), de Sitter and Conformal but now for all dimensions and he showed that \textit{10+1 was the maximal spacetime dimension} compatible with linear Poincaré supersymmetry. He used the above condition of maximal spin 2 in order to obtain this strict bound. For maximal spin one there can be at most 9 space dimensions. It is remarkable that these results do not really presuppose string theory. Werner also proved that the number of spacetime dimensions for de Sitter resp. Conformal supersymmetry is at most seven resp. six.

As we shall see the spectrum of zero mass fields in 11D is very simple, it is composed of a (generalized-)gauge three form, the graviton and a Rarita-Schwinger spinor connection. Werner acknowledges that Jo\"el had conjectured that $\mathcal{N}_4 = 8$  SO(8)-supergravity in 4D might result from compactification to 4D of two of the three possible theories in 10D (types IIA and IIB). The third one (type I) leads to SU(4) supergravity coupled to O(4) super Yang-Mills theory. 

So in the middle of 1977 it became urgent to construct the small slope limit of closed superstring theory. Murray’s and Werner’s maximality results had led Jo\"el to think that $\mathcal{N}_4 = 8$ 4D supergravity was extremely special. So much so that by 1980 Joël had convinced Stephen Hawking that this was probably the ultimate theory. Abdus Salam and Bruno were of the same opinion. This point of view was then embraced by our community at least as a low energy approximation to the full superstring theory. Long before that in 1977 I had asked Jo\"el in the spring and again in the fall what 10D supergravity theory could be. I was interested by the dual model supersymmetries, but he seemed to see that theory as a towering landmark not ripe for our efforts. I thought that the theory was mathematically exceptional, interesting and that it was going to be fruitful but its physical beauty was less obvious and the risk of an end to theoretical physics did not look reasonable to me. In the fall I proposed to him to have a try at the 10D construction and I carried on with his friendly advice. When I later asked Werner why he had not pursued the search of the 11D supergravity action he answered that he did not feel technically prepared for that task. After I made some progress with rather formidable Fierz transformations Jo\"el suggested to me that Eug\`ene should join. 

Then the three of us had a brain storming session and it became obvious that the \textit{11D theory would be easier} to construct because it was polynomial. As a consequence the construction using the Noether deformation method was to be a (big) piece of cake. In retrospect my first attempt at constructing the theory in 10D would probably have also succeeded because it essentially was a scaled up version of the SU(4) 4D endeavour with a single nonpolynomial scalar. But going up in dimension to reconstruct 11D from 10D (oxidation) was an art whereas dimensional reduction was deductive, so why not go to 11D and then downhill?

Among new ideas needed to construct 11D supergravity \cite{CJS1978} let us mention the tensor decomposition of spinor bilinears and the possible choice of a "triangular" moving frame in a \textit{Lorentz symmetry-breaking gauge}. I keep preciously my handwritten notes of our construction as well as Eug\`ene’s notes. We were doing all computations by hand in parallel but with frequent cross-checks. 

* Let us consider first the spectrum of states. The open string vacuum state in 10D is parameterized by an 8 component transverse vector and an 8 component Majorana-Weyl spinor. Closed string theories are “squares“ of open ones.
Hence the 10D vacua of the parity-even IIA superstring theory correspond to a (symmetric) graviton, an (antisymmetric) gauge 2 form and the scalar dilaton trace (the type I part), plus appropriate (antisymmetric) tensors bilinear in the fermions. 
If we were to assume that the eleven dimensional degrees of freedom are a graviton and a spinor-vector gauge connection, this would correspond to 128 fermions but 44 (=128-84) bosons. To restore supersymmetry the \textit{simplest addition} of the missing 84 bosonic d.o.f. would use \textit{an extra 3-form gauge potential $A^{(3)}$}.  

The corresponding degrees of freedom reduce in 10D to the graviton, a gauge vector (1-form) and a scalar, a 3-form and a gauge 2-form plus the spinor-vector connection and a spinor field. The type I 35+28+1=64 dof add to the 8+56=64  of IIA-non-type-I to make 128 bosons. In IIA the number of (on-shell) d.o.f. from antisymmetric tensors bilinear in the fermions can be computed directly. They come from the Clifford algebra generators that carry an odd number of indices (ie 1 or 3) there are 64 of those. In D-brane language these fit the D0, D2 branes and their magnetic duals. In 1987 the reduction from 11D to 10D of a theory of membranes coupled geometrically to a 3-form led indeed to strings coupled to the Kalb-Ramond 2-form and to D2 branes coupled to the mysterious 3-form see \cite{Duff1987} . In string theory the latter form comes from the \textit{bifermion sector} so it is less mysterious but more intricate than in 11D. In \cite{Julia1979} I wrote that this surprising 3-form suggested the existence of \textit{“bubbles”} to source it and to couple it to.

** The second step of a Noether construction is to deform to first order a free action and the “gauge” transformation laws by adding a coupling term proportional to the free current. This is simplest in a 1.5 formalism with moving frame and Lorentz connection fields taken independent. In the supersymmetric case using the \textit{supercovariantization} of the Lorentz connection and of the 4-form field strength F which both appear drags along many fermionic terms. This is a big simplification and helps with the closure of the superalgebra on Bose fields. In the same way as the Lorentz connection is contracted with the commutators of pairs of $\gamma$ matrices that represent the Lorentz generators: $\Gamma^{ab}:=\Gamma^{(2)}$, $F_{abcd}$  (in particular in the Susy variation of the spinor-vector) is to be contracted with the combination  
 $\Delta^{abcde}:=\Gamma^{abcde} +8 \Gamma^{[eab} g^{c]d}$. 
 Here a,b,c,d,e are spacetime indices and $g^{cd}$  is the metric field. One sees that the coupling to the 3-form is \textit{“non-minimal”} ie to the field strength F not to the potential $A^{(3)}$. The torsion-free part of the Lorentz connection is a derivative of the vielbein (the translation gauge field) the latter’s coupling in the transformation law may also be seen as nonminimal.

*** To complete the Noether procedure (which is finite up to diffeomorphism covariantization to be done geometrically) one keeps adding higher order terms of various tensorial types to obtain all quartic terms in the fermions. In so doing we were forced to guess and to add a topological term to the Lagrangian namely  \textit{$A^{(3)} \wedge  F^{(4)} \wedge F ^{(4)} $}. Finally all terms of the variation magically vanish after using a Fierz identity that implies a vanishing combination of bi-Clifford elements: 

$\Gamma^{abcdef} \psi_b \overline{\psi}_c \Gamma_{de} + \Gamma_{de} \psi_b \overline{\psi}_c \Gamma^{abcdef} - 2 \Delta^{abcdf} \psi_b \overline{\psi}_c \Gamma_{d} +2\Gamma_{d} \psi_b \overline{\psi}_c \Delta^{abcdf}  \\
\, \, + 16 \Gamma^{[fab} g^{c]d}  \overline{\psi}_c \Gamma_d \psi_b = 0 $.
 
 The last product exchanges the spinor indices compared to the previous ones see \cite{CJS1978}. A generalized Fierz identity expresses the completeness of a basis of the Clifford algebra or a projection thereof and is related to Pl\" ucker-type identities, it is a central piece of each supergravity construction. The polygamma matrices: the identity, the 11 gamma matrices and all the $\Gamma^{n}$ antisymmetric products of n gamma matrices span a basis of Cliff(11,$\mathbb{C}$). If one chooses to have almost Euclidean signature (one time) the gamma matrices can be taken real. We completed thereby the construction of the 11D supergravity action (without auxiliary fields). The subset of 528 matrices $\Gamma^{(1)}, \Gamma^{(2)}, \Gamma^{(5)}$ is the Lie algebra of $Sp(32,\mathbb{R})$, it contains  so(10,2).

	One must note that one remaining difficulty towards a complete geometrization of our construction is the lack of a manageably infinite set of auxiliary fields that would allow to deduce a fully supersymmetric action. This was discussed in 1980 by Cremmer and Ferrara as well as Brink and Howe for on-shell superspace constructions, for more recent attempts see \cite{Cederwall2013}. We focus here on bosonic symmetries. Sometimes dualities do not seem to require supersymmetry but here probably their ultimate origin is to be searched for in a “fermionic” extension.

\textbf{Fourth Problem}:
	Find a geometric construction of an 11D SUGRA action with infinitely many auxiliary fields. Clarify the role of a possible symmetry like
$OSp(1\mid 32)$ enlarging supersymmetry. Demystify the Fierz magics, the $\Delta$ matrices and other special Clifford combinations
by using Group theory.

\section{$\mathcal{N}=7, 0$
, $D\ge 2$ 
. Affine 
magics. Integrability. }

\bigskip 
\textbf{E groups}: $E_{11-D}$ $\mathcal{N}=7$. It is time to come back down to four (or D) dimensions. Our main breakthrough with Eug\`ene was to imagine a homogeneous space \textit{$R\backslash G$ structure} of the manifold of scalar field values and to guess in D dimensions a (duality) symmetry G (denoted $G^D$) acting on the latter and larger than the expected compact R duality symmetry group. In string theory where it reduces to a discrete group \textit{G is called U-duality}. $G^D$ was deduced by computing/guessing its dimension which luckily was rather transparent because most of the time its Lie algebra is simple and often exceptional. For instance for the D=4 $\mathcal{N}_4 = 8$ supergravity with 70 scalar fields, let us assume that R the most natural invariance subgroup is precisely SU(8) with real dimension 63 it follows that if $G^4$ exists it should have dimension 133=70+63, here comes $E_7$. It turns out that R is the maximal compact subgroup of G (in Lorentzian signature). Consequently $R\backslash G$ is a non-compact symmetric space. For other dimensions from D=10 to D=3 the corresponding group dimensions would be 1, 4, 11, 24, 45, 78, 133 and finally 248. We obtain the family of 8 real Lie groups $G^D=E_{(11-D)}$ split : $\mathbb{R}, \mathbb{R} \times SL(2, \mathbb{R}), SL(2,\mathbb{R}) \times SL(3,\mathbb{R}), SL(5,\mathbb{R}), SO(5,5), E_6$ split, $E_7$ split and $E_8$ split for the $\mathcal{N} =7$ column. Split real form is a mathematical term for "maximally non compact". A ninth group $G^{10B}=SL(2,\mathbb{R})$ arises in 10D IIB supergravity. In 4D R dualities are unitary but in 3D they are orthogonal, in 5D symplectic, beyond 5D octonions do not appear at that naive level.
\bigskip

* The symmetric space structure of the  \textit{scalars} is particularly useful in supergravity. It was very encouraging to confirm our coset assumption by multiple checks like the dimensions of linear representations of G (eg $56=2\times 28$ vectors in 4D) or the action of R on the fermionic fields as \textit{R plays the role of the Spin-Lorentz group} that mediates the couplings of gravity to spinors. 

The coset representative $\mathcal{V}$ in G of the scalar fields resembles the vielbein and its gauge fixed form is a metric. It transforms on the right by multiplication by another (but position independent) element of G and on the left by multiplication by a position dependent arbitrary element of the maximal compact subgroup R of G, sometimes denoted KG. In three dimensions counting up to 248 was done in \cite{CJ1979} and the detailed parametrization of $SO(16)\backslash E_8$ by the scalar fields was worked out in \cite{Julia1983}. The transformations of \textit{spinors} exactly as in General relativity are mediated by the coset fields (ie the internal moving frame) so neither G transformations nor diffeomorphisms act on their indices, only R and Lorentz  
do. This was discovered and analyzed quite exhaustively for D=4 in \cite{CJ1979} other dimensions followed suit. Admittedly the infinite dimensional case in 2D is more difficult, the result of \cite{Julia1981} on the signature of the generalized Killing form was derived first in the D=2 gravitational ($\mathcal{N} = 0$) case but it applies more generally. Another important tool in working out the dimensional reduction to 4D was SO(8) triality. Using internal $\gamma$ matrices we could enlarge the manifest SO(7) toric invariance to the expected SO(8) symmetry in 4D. The latter is included in the SO(10,2) of $OSp(1\mid32)$ and points towards it.
Triality had been envisaged by Werner, Murray and David but it was discovered and implemented in \cite{CJ1979}.

** \textit{Spinning bosonic fields} appear: beyond the scalars (0-forms) there are also (abelian gauge) p-forms. In 4 dimensions the dual of an electric vector potential is its magnetic partner, both mix under dualities so the dimension of the representation $\mathcal{A}$ of G is twice the number of vector fields. In D=2(p+1) dimensions p-form potentials can be self-dual. But quite generally (p+1)-form field strengths $\mathcal{F} = d\mathcal{A}$ are (Hodge) dual to (D-p-1)-forms. There is an action principle associated to any allowed choice of half the set of potentials-and-their-duals. In 4D a Kalb-Ramond 2-form can describe its dual scalar and so on. Off-shell symmetry is maximal when one chooses the minimal degree potential in each pair. To restore the KG gauge invariance we introduced its \textit{compensating non propagating connection}. It cancels the component $\mathcal{Q}$ of $\mathcal{F}^{(1)}_{scal.}$ along Adjoint(KG). 

Please note that  $\mathcal{F}^{(1)}_{scal.} :=d\mathcal{V} \mathcal{V}^{-1}$ is invariant under G and takes values in its Adjoint representation considered as a sum of representations of KG. The orthogonal complement $\mathcal{P}:=\mathcal{F}^{(1)}_{scal.}-\mathcal{Q}$ of $\mathcal{Q}$ is the propagating part of the scalar fields which couple non minimally to the fermions. This coupling resembles the term $\Delta^{(5)} F^{(4)}$ of D=11 in the previous section.  Strictly speaking both the vectors and the propagating part of the scalars do mix spin 1/2 and spin 3/2 particles. A key message is that Physics is done in an inevitably messy fixed gauge but mathematical beauty can and should be resuscitated by \textit{restoring the gauge invariance}. Note however that fixed gauges have their use, for instance the existence of an Iwasawa (triangular) gauge provided automatically by the Noether construction does explain why some fields appear only polynomially despite the nonlinearity of taking the inverse of $\mathcal{V}$. 

***After dimensional reduction the \textit{gravitational sector} can be parametrized most conveniently by using the breaking of the symmetry between internal 
(toric) and remaining spacetime coordinates to choose the moving frame in the friendliest Lorentz gauge \cite{CJ1979}. 
The Lorentz connection can be taken as an independent variable relative to the moving frame at least temporarily in the 1.5 formalism. The Poincar\'e curvature has two parts the Lorentzian curvature and the anholonomy (the combination of torsion and the Lorentz connection that depends only on the frame namely its abelian field strength).

\bigskip

\textbf{D=2.} Clearly a natural extrapolation to D=2 of the $\mathcal{N}=7$  descent suggests \textit{$G^2 = E_9$ also known as (Kac-Moody) affine $E_8$}: the loop group of $E_8$ plus its one dimensional universal central extension) which is infinite dimensional. We learned from several colleagues (Neveu, Schwarz, Thierry-Mieg) that Howard Garland had made this conjecture and that Pierre saw a possible connection to string theory. The theory of representations of affine Kac-Moody algebras was just being developed with inspiration from the bosonic string dual model of Bardakci and Halpern who had actually anticipated affine Kac-Moody representations in 1975 (building up on their 1971 construction). Recall that Kac-Moody theory was born in 1968.

One may convince oneself that after (toric) compactification on $T^k$ the diffeomorphism group of the fiber contains $SL(k,\mathbb{Z})$. After dimensional truncation it becomes $GL(k,\mathbb{R})$ aka $\mathbb{R} \times A_{k-1}$ split: the tangent space symmetry in dimension k \cite{CJ1978}. Quite generally at fixed $\mathcal{N}$ when D decreases (k increases)  in a vertical move the Dynkin diagram of the Lie group $G^D$ grows a line called its \textit{“Gravity leg”} in a regular continuous fashion, see \cite{CJ1979} for $\mathcal{N}= 7$ and D at least 3 and \cite{Julia1980} for D at least 2 and/or lower values of $\mathcal{N}$. This behavior gave more convincing evidence for the emergence of affine Kac-Moody extensions in D=2. The other cases $1\le \mathcal{N} \le 6$ will be discussed in sections \textbf{6} and \textbf{8}. 
Let us note that it is the split real form $GL(D_{max}-D,\mathbb{R})$  that lies at the growing end of a Gravity  leg so when the duality group G is non split we must look for the “split” $A_{D_{max}-D-1}$ inside the noncompact part of its Satake diagram, the plain Dynkin diagram does not suffice for this task. When reducing to dimension 2 the affine (=extended) root of $G^2=G^{3(1)}=G^3\,\hat{}$ appears at the end of the gravity leg. These observations are quite useful to decide whether or not a theory can be oxidized.
 It will be interesting within the $[B=(D-2), F=(8-\mathcal{N})]$
  BF-lattice of theories to compare and contrast the behaviour along fixed $\mathcal {N}$  and variable D columns with that along the lines in the $\mathcal {N}$ direction (with variable number of supersymmetry generators). A line of fixed D is called a \textit{“Susy arm”}. The growth of the Susy arms at fixed D presents a perfect regularity that is visible on the Vogan diagrams of G (B. Julia Princeton slide). In short the R diagram is growing as the main piece of the Vogan diagram; The Satake diagram of G uses the maximal split subCartan of G which can be diagonalized over the reals. The (regular) split subgroups of G ie some of its maximally noncompact subgroups become apparent on the Satake diagram. 
  In the arm direction Vogan diagrams serve the analogous/dual purpose; for regular compact subgroups of G, they use maximal compact Cartan subalgebras, we shall briefly return to the arm/leg symmetry in section{\bf 7}.

We began this chapter with the discrete arithmetic modular group $SL(2,\mathbb{Z})$ as the quantum modular invariance and just met it as the  symmetry of a two torus during classical compactification. In 4D the dualities of quantum string theory and of quantum field theories are typically the intersection of G (real Lie group) and $Sp(2n,\mathbb{Z})$ with n the number of vectors: again an arithmetic group, see the contribution of Chris Hull.   
\bigskip

\textbf{The pure gravity case.} In 1979 just after I finished presenting our last results with Eug\`ene at a summer conference in Trieste Stephen speeded towards me in his wheelchair followed by Gary Gibbons and he remarked that the dimensional reduction of \textit{$\mathcal{N}_4 =0$ pure 4D general relativity} to 3 dimensions looked somewhat similar to the $\mathcal{N}_4 = 8$ theory reduced to 3D with $SL(2,\mathbb{R})$ the Ehlers duality symmetry replacing $E_8$. In other words the very short gravity leg for \textit{
$\mathcal{N}=0$} resembles the long one for $\mathcal{N}=7$ \footnote{But in his euclidean framework our SO(2) was replaced by a non-compact SO(1,1). }
Geroch had discovered in 1972 that in the presence of two commuting Killing vector fields and provided two constants vanished the SL(2,$\mathbb{R}$) Ehlers symmetry of each preserved the other so Geroch generated a (his) mysteriously infinite dimensional group from all of these 3 dimensional Ehlers subgroups. Even Stephen and Gary did not know exactly what it was but they agreed with me that its infinite dimensionality was to be understood. The original Ehlers $SO(2)$ inside $SL(2,\mathbb{R})$ is a rotation between ordinary mass (electric or fifth dimension) and magnetic mass (or NUT-charge) (sixth dimension). I believe I taught this six dimensional interpretation to D. Olive in 1977: the corresponding two extra $\gamma$ (or rather $\sigma$) matrices differ by the $\Gamma^{56}$ factor, which in four dimensional notation is $i\gamma ^5$ as we remarked above (hence my interpretation of duality). Following up on the above remark on the realization of dilations by 5-6 hyperbolic rotations, it should be interesting to break simultaneously scale invariance and chirality (and electric-magnetic duality) by working over the complex numbers.

\textbf{Geroch group as affine SL(2).} In the winter of 1981 I was invited by Murray to visit Caltech where I struggled with the General relativity literature in order to make sense of the Geroch group as an affine Kac-Moody group. Towards the end of my stay I got a sweet and sour comment from Christopher Cosgrove to whom I presented my conjectures (I had not fully checked the completeness of the set of relations for the algebra presentation) that one of the Ehlers groups and the gravity leg SL(2,$\mathbb{R}$) symmetry formed the canonical presentation of the Kac-Moody algebra \textit{affine SL(2,$\mathbb{R}$)} including the central charge and mutatis mutandis for Einstein-Maxwell vacua or the $E_9$ case. Christopher recognized (partly) the root set from unpublished work of W. Kinnersley et al. but he claimed too optimistically that my conjecture for the Geroch group had already been established which it had not been. In 1981 there had neither been any precise mention by general relativists nor clear evidence of affine symmetry for pure gravity or Einstein Maxwell vacua  (Victor Kac's book appeared in 1983).  I gave a talk in Baltimore \cite{Julia1981} where I presented my results. I explained there how the generalized Killing form could be used to define a "maximal compact" subgroup $KG^2$ of the duality group $G^2$: as in the finite dimensional case (eg $G^4=SU(1,1)$ above) one may select the negative-Killing-normed generators and call them “compact”. Furthermore I also studied the action of the Geroch group $G^2$ on the conformal factor of the 2D spacetime metric and discovered the central extension where it should be. A special feature of 2D is that one cannot scale away a factor in front of the scalar curvature as one does in other dimensions. This is the reason why a central charge may appear there.

I missed the fact that some of my solution generating transformations were not symplectic and as a result I missed the invention of Quantum groups or at least of \textit{Lie-Poisson groups}! Quantum groups were mathematically defined in 1985 by Michio Jimbo and Vladimir Drinfeld (see also his letter in 1983) but examples had been worked out and studied by physicists of the Leningrad school before (Kulish, Sklyanin, Semenov-Tian-Chanski, Reshetikhin… 1979-1982)! This impressive school was started by L.D. Faddeev in 1978! It was actually discovered soon after \cite{Julia1981} that some Geroch solution generating transformations were not “canonical” \cite{Davies1982}. 
I did visit the Leningrad group at the end of 1982 and presented my results there as well as at the I.M. Gel’fand seminar in Moscow. 
Fortunately some of my observations remained useful in particular the identification of a finite generating set including in the gravitational case the central charge see \cite{BM1987}. Breitenlohner and Maison did check the commutation relations and the symmetry group, they related them to Dieter Maison’s linear system \cite{Maison1978}. The discussion of a classical Lie-Poisson structure and of the quantum group obtained thereafter will not be covered here, it seems to still suffer from analytical difficulties with asymptotics. The so-called nonlocal charges Lie-Poisson action was discussed by D. Korotkin and H. Samtleben in 1997-8. The precise interplay of the affine Kac-Moody solution generating symmetries and Lie-Poisson ("Dressing") results deserves a full mathematical resolution in the spirit of M. Semenov-Tian-Chanski’s construction of a \textit{“classical” double} in 1985. 

\textbf{Integrability?} The 4D \textit{twisted self-duality} equation for coupled vector and scalar matter \cite{CJ1978}  $\Omega \mathcal{V} * \mathcal{F}^{(2)}=\mathcal{V} \mathcal{F}^{(2)} $  generalizes to all higher forms of complementary degrees in \textit{any} dimension \cite{CJLPII1998} as we shall see in the next section. The square of the invariant $\Omega$ must be $\pm1$ depending on the signature of spacetime. An important goal is to reexpress the full Einstein equations in a similar form. An instanton formulation of "all Einstein spaces" in D=4 with any value of the cosmological constant was discovered by Belavin and Burlankov in \cite{BB1976}. It is in fact also a \textit{twisted self-duality} equation and reads: $^*\, _*R^{(2)}=R^{(2)}$ for the curvature 2 form valued in the Lorentz algebra provided we set the torsion to zero. The first Hodge star-dualisation acts on Lorentz indices and plays the role of the R invariant operator $\Omega$ above, the second one simply acts on the 2-form as usual. Note that Atiyah et al. \cite{AHS1977} essentially rediscovered this and proved that the bundle of self-dual two forms on a Riemannian 4-manifold $\mathcal{M}_4$ has self-dual curvature if and only if the base space is Einstein (in the torsion free case). They then used the Penrose twistor equation to identify complex structures on the twistor space of projective chiral Weyl spinors with conformal structures on $\mathcal{M}_4$ such that the Weyl curvature tensor is self-dual. We may mention that C. LeBrun encountered troublesome torsion in his work on general twistor solutions of gravity. The present analysis must now be extended to include the matter fields of pure SUGRAS still in 4D.

Even in D=4 it is not clear how to go beyond instantons as one looses the original twistor interpretation. Tentative generalizations for integrating the full Yang-Mills (and Einstein) equations in 4D are reviewed in \cite{Witten1986} but one is led to null geodesics and the dictionary remains incomplete. One general tool that allows to start the generalization is to double the dimension of space time and to use formal (finite order) neighbourhoods.  
\textit{Twisted self-duality} means that all the second order equations of motion (and Bianchi identities) can be rewritten as self-duality (more generally auto-B\"acklund) first order equations for a doubled set of fields. A true \textit{unification} between the matter and gravitational sectors \textit{should implement this on both sectors}. 
The same moving frame idea applies in both internal sigma model space $\mathcal{V}$ and in spacetime in the Geroch case. 2D self-duality is trivially realized on the harmonic dilaton field $\rho=\sqrt{det g_{ij}} >0$ that appears upon reduction in the 2 dimensional internal part $g_{ij}$ of the 4D general relativistic metric. The coordinates (angles) of the "Virasoro" (gravity) generators $L_0$ and $L_{-1}$ are the non propagating "matter" fields $log\rho$ and its harmonic conjugate $\tilde{\rho}$ in off diagonal combinations  \cite{JN1996}.
It seems very doable to express the group theoretical meaning of not only the harmonic dilaton and  its dual  but also of the conformal field $\sigma$ and its dual where $exp 2 \sigma$ is the conformal factor of the 2D spacetime metric.
The appearance of two different scattering parameters for instance in Maison's treatment ought  to be deduced from the semidirect product structure of the symmetry group that contains now the Witt ("Virasoro") algebra.
See also a “dressing method” approach in \cite{BJ1999}. 
\bigskip

A main criterion of integrability of a classical non linear differential problem is the existence of a linear system that is compatible if and only if the nonlinear equations are obeyed. In 2D field theories a principal source of those is the Yang-Mills instanton equations reduced to 2D in various ways. The spectral parameter that appears there is the modulus of anti-self-dual null planes encoding the vanishing of some Yang-Mills curvature components.  

Finally let us now return to the fully 4D problem: a linear system for the 4D vacuum Einstein equations (without spectral parameter) was found in \cite{Julia1982}. The linearized gravitino equation of $\mathcal{N}_4=1$ are by construction linear and compatible or Cartan integrable if the full 4D vacuum Einstein equations are satisfied. Although this linear system has no spectral parameter could we add one? 
Again this should be answerable probably in the negative but the computation might be instructive.


\textbf{Fifth Problem:}
Characterize 4D Gravity's special integrability properties in view of the preceding remarks and similarly for SUGRA theories.

\section{$E_{10}$, chaos. Borcherds. $E_{11}$. SUGRA Triangle.} 

{\bf D=1 chaotic (hyperbolic) Magics.}

In the previous section we saw the emergence of integrability in 2D together with the emergence of affine Kac-Moody symmetries. What we called affine KM magics deserves to be called "Integrable Magics" more generally. The Gravity leg argument will lead in 1D in the case of maximal supergravity to $G^1=E_{10}$ where now $E_8$ is overextended to
$E_8 \hat{\,}\, \hat{\,}$. In 1982 I was invited by I. Singer to a conference in Chicago where I met Igor Frenkel who resonated and was very helpful. The Magic square of Tits and Freudenthal had fascinated him, so my SUGRA triangle \cite{Julia1980} attracted his attention\footnote{Alexander Kirillov also kindly mentioned his special interest to me.}. Among Kac-Moody algebras with one negative normed root direction the so called \textit{hyperbolic Kac-Moody algebras} are simplest and very constrained. For instance beyond rank 2 there is only a finite number of them: 4 in maximal rank=10 out of which $E_{10}$ and 2 more hyperbolics occur in string theory. I included immediately my improved understanding into a preprint \cite{Julia1982-5} in which I conjectured $G^1$ to be the hyperbolic symmetry \textit{$E_{10}$} resp. \textit{$\mathcal{F}_3$}  of \cite{FF1983} in the $\mathcal{N}=7$ maximal supergravity case resp. in the minimal pure gravity column or equivalently the $\mathcal{N}=1$ supergravity case. The corresponding chaotic behavior was checked 19 years later \cite{DHJN2001} for {$\mathcal{F}_3$.	
The Dynkin diagram of $\mathcal{F}_3$  is again the overextension $A_1  \hat{\,}\, \hat{\,}$ of $A_1$. It is a hyperbolic Kac-Moody algebra of rank 3 and was being introduced for number theoretical purposes by Alex Feingold and Igor at the time of my visit. I also conjectured that similar overextensions take place for other theories. $G^1=G^3  \hat{\,}\, \hat{\,}$. This notion of overextension \cite{Julia1982-5} is now standard in group theory. 

Around Christmas 2000 Thibault presented his work with Marc Henneaux on hyperbolic billiards and BKL chaos in 10D string models. After his seminar at IHP-CEB\footnote{ Following my proposal and lobbying the “Emile Borel thematic Center” was added in 1991 to Institut Henri Poincar\'e which  had been created in 1929 by George Birkhoff and Borel to open up the french theoretical Physics community!}
on the $\mathcal{N}=7$ billiard I speeded towards him as Stephen had done and I mentioned Jacques Monod's unification of E. Coli and Elephant: ”What is true of supergravities is true of gravity”, a simultaneous study of different values of $\mathcal{N}$ is very useful as we saw already; shortly thereafter we found the Einstein billiard \cite{DHJN2001}. The non-compact symmetric spaces have constant negative curvature, this is the cause of chaos in a situation where symmetries are plentiful and formal integrability remains in the background.

{\bf Borcherds supersymmetry, twisted self-duality. }

Beyond general Kac-Moody (super)algebras there are Borcherds-Kac-Moody (super)algebras which are sometimes more manageable and related to string theory. 
Already in 11D one would like to unify the 3-form and its dual 6-form. This was first done in 1997 see \cite{CJLPII1998} where we could combine all generalized gauge invariances for the propagating Lagrangian form fields and their duals into a finite dimensional superalgebra with coefficients (angles) in the exterior algebra of spacetime. This larger symmetry wa s called \text {V-duality} as it contains the U-duality G. Let us consider the simplest example of 11D SUGRA to do step one: double the fields and construct the generators and step two: multiply k-form fields of the $\mathbb{Z}$-graded exterior algebra by degree (-)k generators of  an appropriate $\mathbb{Z}$ and hence $\mathbb{Z}\slash 2\mathbb{Z}$ graded duality superalgebra. For scalar fields (the G symmetries considered up to now) we are used to non linear couplings and non abelian dualities; for higher abelian gauge-potentials their duals typically transform nonlinearly by incorporating lower degree potentials.  Let us call E the odd generator associated to the 3-form $A^{(3)}$ and M the "bosonic" generator associated to the dual 6-form $B^{(6)}$. In our example the bosonic exponent is $AE$ or $BM)$ for the group. If we take all (anti-)commutators to vanish but $\{E,E\}=-M$ the ensuing non-linear combinations simplify enormously: the equations of motion and Bianchi identities obey the universal \textit{general twisted self-duality equation} which reads $*\mathcal{G}=S\mathcal{G}$ wiith a supergroup potential field $\mathcal{W}:=\mathcal{V} e^{AE}e^{BM}$ and its field strength $\mathcal{G} = d\mathcal{W} \mathcal{W}^{-1}$; the invariant operator $S$ has square $\pm 1$ depending on the signature. We kept $\mathcal{V}$ even though it is trivial in 11D but in general our familiar scalar fields appear there. 
V-duality discovered in \cite{CJLPII1998} was recognized as Borcherds (super)symmetry in  \cite{HJP2002}. This is still to be streamlined and better understood but this is Algebraic Magics.
We shall only mention the example of type IIB which has a purely bosonic (only even degree potentials appear) algebra: it is a truncated positive grade part of the rank 2 Borcherds algebra studied by Slansky in 1993. Amazingly (but for other superalgebras a simple correction is mysteriously required) its (Borcherds-Kac-Moody-)Cartan matrix is the middle degree cohomology intersection form on a corresponding Del Pezzo surface which in the IIB case is the K\"ahler-Einstein $CP^1\times CP^1$ \cite{HJP2002}. That same rational surface blown up in $k\le 7$ points in general position is related to the Weyl group of $E_{(k+1)}$! 

Let us insist that the V-duality is only a truncation of the Borel (positive degree) part of the Borcherds superalgebra we introduced. Contrary to the U-duality it has not yet been extended to a full Borcherds realization. I believe that the main obstacle is to invent a way to mix differential forms and multivectors (negative degree forms). Recent ideas might be brought to bear on this for instance \cite{Kosmann2004}, see below. 

{\bf From $E_{10}$ to $E_{11}$ and embedding tensor.}

In (infinite dimensional) Kac-Moody theory the most important algebras are the affine ones. The next generalization are hyperbolic KM among Lorentzian KM. More general ones did not attract much attention yet outside physics. 
The duality group $G^0$ after full compactification to D=0 is naturally expected to be  "$E_{11}$". The main argument for choosing $E_{11}$  is that for all D its diagram contains exactly  the factor diagrams of the product of the duality group $E_{11-D}$ by $SL(D,\mathbb{R})$ separated by a middle Cartan generator.
In 2 lectures in Cargèse in 1997 I explained that the field representations of these two factors were defined by highest weights with a nice regular pattern. The vectors form a fundamental representation of each factor with highest weight at the ends joining the E and the A subdiagrams, the scalars belong to the adjoint of E times the trivial representation of A, the two forms use two other vertices... In Fig. 3 of \cite{Julia1997} the choice of $E_{11}$ and of its decompositions is clearly visible. Subsequently Peter West decomposed this way the adjoint representation and showed that a dual graviton field appears automatically therein.

The nonlinear gauging of the available vector fields turns out to be extremely rich and mysterious. The first breakthrough was due to B. de Wit and H. Nicolai, see the contribution of Hermann Nicolai or \cite{Trigiante}. 
The first structure that emerges in this search is the embedding tensor that relates the vector potentials to the generators of U duality. This leads to the Tensor Hierarchy Algebras THA \cite{Palmkvist2012}.  One can recognize the embedding tensor inside THA and Borcherds but also inside $E_{11}$.
In fact in \cite{HJL2010} we constructed the exceptional duality algebras (for all D) by tensoring the $E_k$ generators with the exterior algebra on the base manifold but this time \textit{keeping only spacetime scalar invariants}. So we extracted the Borcherds structure from $E_{11}$ which contains more tensors than p-forms. One can recover also the non propagating p-forms in this way.
 A modern and systematic approach to gaugings ought to use the BV formalism, see for instance \cite{BJulia2017}.

   \textbf{SUGRA triangle. }                        
In July 1980 in \cite{Julia1980} $E_9$ was discussed as well as some of its non split analogues in D=2 for lower $\mathcal{N}$.
It was shown that the surprise appearance of the Kac-Moody affine extension upon descending from 3D to 2D held for many $\mathcal{N}$’s. 
More generally all "pure" supergravities were organized in an embryonic triangle see section {\bf 8}.
It was also noticed then that for two duality groups associated to the pair $(B,F)$
$(D-2=9-k:=B, 8 - \mathcal{N}:=F)$ 
and its symmetric image across the diagonal  (B'=F, F’=B)  \textit{their  complexifications $G^\mathbb{C}$ were the same}. But the set of supergravities that reduce to pure supergravities in 4D is actually dissymmetrical across the diagonal I called it the (first) SUGRA Magic triangle although it looks more like a trapezoid. His horizontal side has (B,F) from (1,1) to (1,8) and the vertical one from (1,1) to (9,1) (resp. (8,1) for a chiral IIB version), this problem is also encountered in the “Magic pyramid” of Imperial College \cite{Borsten2019} and references therein.
This was still a wild conjecture comforted by the growth of the Gravity leg. 
It contains except for its smallest (corner) group, see the Table in the Appendix, a real form of the 3$\times$3 subMagic square of Tits and Freudenthal. For a review over the complex numbers see \cite{Cvitanovic2020}. But Tits has shown that no real version of his constructions can lead to all our supergravity real forms. Let us note that our duality results invove real forms of Lie groups and are finer than most of the complex litterature we are associating them to. But finer means also more sophisticated and more difficult.  The appendix contains the table with the SUGRA triangle  the next challenge refers to it.  The SPLIT triangle is presented in the next section and also drawn in the table.
 
 \textbf{Sixth Problem:}
Is there an 8th column of theories $\mathcal{N}= 8$ through $\mathcal{N}_4 = 16$? In 4D it would have by BF-reflection ($\Sigma$) symmetry a large invariance algebra which could be a real form of the complexification of the above semi-direct product of the Witt algebra by the non-split affine $E_7^{(1)}$.

\section{BF symmetry, SPLIT+Cvitanovic Triangles}

\textbf{SPLIT triangle $d-n\le 3$, Metaduality D-2 $<->$ 8-$\mathcal{N}$.}

The SUGRA triangle admits a partial symmetry (Metaduality) let us call it $\Sigma$ across the diagonal. 19 years later the original request for supersymmetry was discarded and the series of split $E_{n+1}$ non-compact symmetric spaces was chosen to replace the real forms of the previous D=3 arm \cite{CJLP1999}.
In the Table below I superposed to the first Magic SUGRA-triangle the second Magic SPLIT-triangle whose columns were constructed by oxidizing each of the bosonic sectors based on the new split $E$’s  from d=3 to $d_{max}\sim n+3$. Surprisingly by enforcing the identity of the two sides of the second triangle but despite its asymmetric construction the symmetry across the diagonal became perfect inside too as we see in the lower left part of the Table. In particular one gets the full $3\times 3$ part of the Magic square.
 $\Sigma$ is a perfect symmetry of the resulting second triangle of (split) real forms for bosonic theories whose groups are also simply laced but for generic n we do not have any supersymmetry anymore.  We shall discuss the missing $F_4$ and $C_3$ of the $4\times 4$ Magic momentarily.

Now the meta-symmetry $\Sigma$ suggests that interesting features of Susy arms could be dual to properties of Gravity legs. 
In the SUGRA triangle the Susy arms still grow stepwise with $\mathcal{N}$ at least their R=KG subgroups. The arms for D=3 and 4 are perfectly regular. For R the arm grows linearly but more interesting for G also, its "compact line"  has a the regular growth similar to that of the Gravity leg. Vogan diagrams encode the maximal compact Cartan subalgebras which may not be unique and their associated "regular" subalgebras. For D=5 and 6 with two instances each it is more subtle, this is unpublished work of the author presented in seminars a few years ago. 
One may wonder why $\mathcal{N} _4$ plays such a role. Obviously in D=3 odd $\mathcal{N}_3$ cases cannot be oxidized to 4D, they have disappeared because only even ones can. But Bernard, A. Tollsten and Hermann Nicolai constructed in 1992 \cite{WTN1992} non simply laced duality groups G  for instance $F_4$  for $\mathcal{N} _3=9$ or $C_3$  for $\mathcal{N} _3=5$ and also $G= A_2$ with $\mathcal{N} _3=3$. 
I remarked in \cite{Julia2014} that to extend the $\mathcal{N}, D$ exchange symmetry $\Sigma$ to those cases would require half odd D, e.g. for the $F_4$  symmetry partner D’=$11/2$ which could be a quantum dimension, it would lie just next to $E_6$ as in the magic square similarly for $C_3$  D’=$15/2$ and for $A_2$  D’=$17/2$.

Let us repeat that even if supersymmetry is escaping our senses its mathematical power is obvious. The exceptional duality symmetries of the supergravity triangles are bosonic but probaby deeply related to supersymmetries and fermions via the $R\backslash G$ coset frame and many related aspects. On the other hand $\Sigma$ is enhanced when one relaxes the supersymmetry constraint!

One may also think that the metaduality $\Sigma$ might be related to colour-kinematic duality, this remains to be seen but an amusing application of the latter to the symmetry of their pyramids has been discovered by Mike Duff's group \cite{Borsten2019}
The parallel between $\mathcal{N} = 0$ and $\mathcal{N} = 7$ and more generally between all $\mathcal{N}$ values is rather important because it anchors firmly all the mathematical games and physical speculations of supergravity/superstring theory to the large body of knowledge of pure gravity and also of Einstein-Maxwell theory both being physical theories. 

\bigskip

\textbf{Cvitanovic' triangle.} 

 A big part of a second type of Magic triangles allowing non simply laced groups and containing the full magic square had been discovered in an Oxford preprint of 1977 by Predrag Cvitanovic \cite{Cvitanovic2020}, in full symmetric form in \cite{Rumelhart1995} and much clarified in \cite{DelGross2002} and subsequent works. Applications of both types of triangles are plentiful from Supergravities to Del Pezzo surfaces and multiplicative discrete Painlev\'e equations for the first type resp. from Lie Invariant theory to several types of Painlev\'e equations (and maybe the Kervaire invariant one problem: $D_4$ / $G_2$ families, B. Julia unpublished) for the Cvitanovic triangle. Once more the "exceptionality" of $D_4$ reveals itself and Howe dual pairs are at work.  


\textbf{Seventh Problem:}
We learned above that 6D is very rich in particular with the ultrahyperbolic signature (3,3). That signature is very rarely used in physics because it does not allow the causality we need to predict and even to think.  Algebraic geometry does exceptionally like in the DelPezzo case provide the nice hyperbolic signature. Is there a way to recover some predictive power and time ordering when our environment is not that kind and provides a higher dimensional time? 

One idea but we are charting very unfamiliar ground is to find a "causal sector" where predictions remain possible. For instance a U(1) gauge symmetry mixing a circle of time directions  may preserve a radial time. This would correspond to the signature (2,3). The case of signature (3,3) would require the compactification of a 2-sphere of times...
Some discussion of the dimensionality of spacetime can be found in \cite{Tegmark}.

\section{Appendix: SUGRA and SPLIT triangles} 

The trapezoid of supergravity dualities G discussed above is reproduced in the upper right hand corner with coordinates 8-$\mathcal{N}$ and D-2 .
Its split completion \cite{CJLP1999} is coordinatized with 8-n and d-2  in the lower left corner. WARNING in order to save space in the Table whenever G and R contained a common compact factor it was ruthlessly eliminated from G. But this hides the equality of the complexifications of SU(1,1)$\times$SU(4) and SU(2)$\times$SO(5,1)...!
\bigskip 

\begin{center}
\begin{tabular}{|c|c|c|c|c|c|c|c|c|}\hline
$_{8-n\backslash}^{\backslash8-\mathcal{N}}$&$_{1}\backslash^{8}$&$_{2}\backslash^{7}$&$_{3}\backslash^{6}$&$_{4}\backslash^{5}$&$_{5}\backslash^{4}$
&$_{6}\backslash^{3}$&
$_{7}\backslash^{2}$&$_{8}\backslash^{1}$ \\ \hline 
$\backslash ^{-1}$ & $\mathcal{F}_3$  &$<$ - &-&-&-&-&-$>$&$E_{10} G^{3 hyp}$\\ \hline
$\backslash ^0$ &$A_1^{(1)}$& $<$ -&-&-&-&-&- $>$&$ E_9 G^{3 aff}  $ \\ \hline
$\backslash ^1$&$A_1$&$A_1$&$SU(2,1)$&$SU(4,1)$&$SO(8,2)$&$E_6$:EIII&$E_7$:EII&$E_8=G^3$\\ \hline
$_9\backslash^2$ &$_1\backslash^1$ & 1& 1&1&SU(1,1)&SU(5,1)&SO*(12)& $E_7$\\ \hline
$_8\backslash^3$ &$\R$ or $A_1$ &  &&&&&SU*(6)&$E_6$ \\ \hline  
$_7\backslash^4$ & $\R\times A_1$ & $\R$ & & &&&SO(5,1)&$D_5$ \\ \hline
$_6\backslash^5$ & $A_1\times A_2$ & $\R \times A_1$ & $A_1$ & &&&&$A_4$\\ \hline
$_5\backslash^6$ & $A_4$ & $\R\times A_2$ & $\R \times A_1$ & $\R$ & 1 & & &$A_1\times A_2$\\ \hline
$_4\backslash^7$ & $D_5$ & $ A_1\times A_3$ & $\R \times A_1^2$
        & $ \R ^2$ & $\R$ & & &$\R\times A_1$   \\ \hline
$_3\backslash^8$ & $E_6$ & $A_5$ & $A_2^2$ & $
       \R\times A_1^2$ & $\R \times A_1$ & $A_1$ & &$\R$ or $A_1$ \\ \hline
$_2\backslash^9$ & $E_7$ & $D_6$ & $A_5$ & $A_1\times A_3$ &
        $\R \times A_2$ & $\R \times A_1$ & $\R$ &$_1\backslash^1$\\ \hline
$_1\backslash$ & $G^3=E_8$ & $E_7$ & $E_6$ & $D_5$ & $A_4$ & $A_1\times A_2$ & $\R \times
A_1$ & $\R$ or $A_1$\\ \hline
$_{d-2\backslash}^{\backslash D-2}$&&&&&&&&\\ \hline
\end{tabular}
\end{center}
\bigskip
\centerline{{\bf Table:} Simplified Duality groups G. NE (D,$\mathcal{N}$) SUGRA and SW (d,n) SPLIT triangles }

\bigskip

{\bf Acknowledgements.}
Beyond the mathematicians and physicists quoted in the text important conversations with Jean-Benoit Bost, Michel Dubois-Violette, Feza G\" ursey, Yvette Kosmann-Schwarzbach, Dimitry Leites, John McKay, Yuri Manin, Louis Michel, Olivier Schiffmann, Adam Schwimmer, 

\noindent Graeme Segal, Ramamurti Shankar, Peter Slodowy, Dennis Sullivan, Gerardus 't Hooft, Charles Thorn, Cumrun Vafa, Claude Viallet and Barton Zwiebach are also gratefully remembered.

\end{document}